\documentclass[12pt]{article}
\usepackage{hyperref}
\usepackage{amsmath}
\usepackage{comment}
\usepackage{enumerate}
\usepackage{cite}
\numberwithin{equation}{section}
\title{Rotating Einstein--Maxwell black holes\\ in odd dimensions}
\author{Rhucha Deshpande\footnote{rdeshpande@albany.edu}\  \   and Oleg Lunin\footnote{olunin@albany.edu}}
\date{}
\begin{document}
\def\be{\begin{equation}}
\def\bea{\begin{eqnarray}}
\def\ee{\end{equation}}
\def\eea{\end{eqnarray}}
\def\d{\partial}
\def\eps{\varepsilon}
\def\la{\lambda}
\def\b{\bigskip}
\def\nn{\nonumber \\}
\def\p{\partial}
\def\t{\tilde}
\def\h{{1\over 2}}
\def\r{{\rho}}
\def\qb{\bar q}
\def\be{\begin{equation}}
\def\bea{\begin{eqnarray}}
\def\ee{\end{equation}}
\def\eea{\end{eqnarray}}
\def\b{\bigskip}
\def\u{\uparrow}

\maketitle

\begin{center}
\ \vskip -1.2cm
{\em  Department of Physics,\\
University at Albany (SUNY),\\
1400 Washington Avenue,\\
Albany, NY 12222, USA
 }
 \end{center}

\vskip 0.08cm
\begin{abstract}
\noindent 
To construct higher--dimensional counterparts of the Kerr--Newman black holes, we consider Einstein's equations sourced by a vector field and a negative cosmological constant. In contrast to the four--dimensional case, the Maxwell's equations are modified by sources generated by  topological Chern--Simons couplings, the situation already encountered in the minimal five dimensional supergravity. After constructing explicit geometries in all odd dimensions, we demonstrate that the Klein--Gordon equation on the new backgrounds is fully separable.

\end{abstract}
\newpage
\tableofcontents


\section{Introduction}

Over the last several decades there has been a remarkable progress in understanding of black holes in higher dimensions and in applying them to studies of quantum gravity \cite{BHapQGw1,BHapQGw2,BHapQGw3,BHapQGw4} and strongly coupled systems \cite{maldW1,maldW2,maldW3,maldW4,AdCFbhW1,AdCFbhW2,AdCFbhW3}. In the absence of rotation, one can easily generalize the Reissner--Nordstrom black hole to arbitrary dimensions by adding charges to the Schwarzschild--Tangherlini solution. In the rotating case, even the neutral solutions turned out to be rather nontrivial, and they are given by the Myers--Perry geometry \cite{MP} and its generalization to the system with cosmological constant \cite{GLPPw1,GLPPw2}. Charged versions of the Myers--Perry geometry can be constructed using the U--duality techniques developed in 
\cite{SenW1,SenW2,SenW3,SenW4,SenW5,SenW6}: they involve adding a torus to the system and performing various string dualities. In the presence of the cosmological constant, the procedure of adding charges is much less straightforward, and it had been applied only on the case--by--case basis by embedding systems into gauged supergravities\footnote{Examples of charged solutions in gauged supergravities in various dimensions include \cite{AdCFbhW1,AdCFbhW2,AdCFbhW3,GSR4slnW1,GSR4slnW2,GSR4slnW3,GSR5slnW1,GSR5slnW2,GSR5slnW3,GSR5slnW4,
GSR5slnW5,GSR5slnW6,GSR6slnW1,GSR6slnW2,GSR7slnW1,
GSR7slnW2,10auth}. Beyond gauged supergravities, very few charged asymptically--AdS geometries have been constructed \cite{ByndGauSugraW1,ByndGauSugraW2,ByndGauSugraW3}.}. Both the duality techniques and gauged supergravities produce solutions which have various scalars in addition to gauge fields, so even in four dimensions the geometries do not reduce to the Kerr-Newman (KN) black holes.

Given the importance of the Kerr-Newman geometry, it is natural to ask for a higher dimensional generalization of this solution supported only by a gauge field. Unfortunately, numerous attempts to find rotating solutions of the Einstein--Maxwell theory beyond four dimensions have failed \cite{KNdFailW1,KNdFailW2,KNdFailW3,KNdFailW4,KNdFailW5}, but in article \cite{BH5d} a very close cousin of the KN solution was constructed in five dimensions. Specifically, the authors of \cite{BH5d} solved the Einstein's equations with Maxwell sources, but they modified the equations for the gauge fields by adding a topological term to the action. We will extend this logic to all odd dimensions by constructing rotating solutions of Einstein's equation with Maxwell sources in the presence of a negative cosmological constant. To satisfy equation of motion for the gauge field, we will add some topological terms to  the action, and beyond five dimensions such terms will involve non--dynamical auxiliary fields. The resulting solutions are probably the closest analytical counterpart of the KN black holes in higher dimensions. In this article we focus on odd--dimensional solutions with equal rotations since in this case the symmetries of the problem lead to a relatively simple ansatz where most metric components are functions of only one coordinate. In section \ref{SecSoln} we will comment on challenges associated with extending this ansatz to non--equal rotations. 

One of the most remarkable properties of the Myers--Perry and Gibbons--Lu--Page--Pope geometries \cite{MP,GLPPw1,GLPPw2} is full integrability of dynamical equations for all fields on these backgrounds. As shown in \cite{KytMpOrgW1,KytMpOrgW2,KytMpOrgW3,KytMpOrgW4,KytMpOrgW5}, these geometries admit a family of Killing--Yano tensors (KYT) which guarantee separability of equations for scalars and spinors\footnote{In \cite{ChL} the construction of KYTs was extended to charged geometries obtained from the Myers--Perry solutions by applying string dualities.}. The equations for vector fields and higher forms were also shown to be separable \cite{MPmaxwOL,MPmaxwFKw1,MPmaxwFKw2,MPmaxwFKw3,MPforms}. 
This raises a question about persistence of such integrable structures for the charged geometries constructed in section \ref{SecSoln} of our article. In section \ref{SecKG} we demonstrate that the equation for a charged scalar field remains fully separable in all dimensions, while the Killing--Yano tensors seem to disappear\footnote{Such disappearance has been already noticed for the five--dimensional CCLP black holes constructed in \cite{BH5d}, although in that case one could define a KYT with torsion \cite{KubizKYT5}. Unfortunately, the procedure for introduction of torsion from the Chern--Simons coupling seems to be specific to five dimensions.}.

This paper has the following organization. In section \ref{SecSoln} we construct charged rotating geometries in all odd dimensions by proposing an ansatz for the metric and solving the Einstein's equations sourced by a gauge field. We also propose the Chern--Simons coupling to additional auxiliary fields which ensure that equations for the gauge field are satisfied as well. This Chern--Simons coupling is similar to the one encountered in the minimal five--dimensional supergravity in \cite{BH5d}, although in that case no fields in addition to the vector were required. In section \ref{SecProperty}, we analyze various physical properties of the new solutions, such as their entropies, temperatures, and charges. In section \ref{SecKG} we study dynamical excitations of the new solutions and show that the charged Klein--Gordon equation is fully separable in all odd dimensions. Some technical details are presented in appendices. Specifically, Appendix \ref{App5d} reviews the Chong--Cvetic--Lu--Pope (CCLP) solution of the minimal five--dimensional dimensional supergravity which serves as an inspiration for our construction. In the same appendix we also demonstrate full separability of the charged Klein--Gordon equation in the CCLP geometry to connect to our discussion in section \ref{SecKG}. Appendix \ref{AppFrames} reviews the neutral rotating black holes in all odd dimensions and special frames that play an important role in ensuring separability of various dynamical equations. In the same appendix we also take a singular limit of the known frames to cover the case of equal rotations relevant for sections \ref{SecSoln} and \ref{SecKG}. Appendix \ref{AppKappa} outlines the general procedure for computing surface gravity of higher dimensional black holes and applies it to our new solutions. Appendix \ref{AppAfield} provides some technical details supporting the analysis presented in section \ref{SecKG}.

\section{Generalization of the KN--AdS black hole}
\label{SecSoln}

The main goal of this article is to generalize the Kerr-Newman geometry to higher dimensions, i.e., to construct solutions of the Einstein's equations sourced only by a gauge field. Previous attempts to solve 
this problem have failed \cite{KNdFailW1,KNdFailW2,KNdFailW3,KNdFailW4,KNdFailW5} due to a clash between the Einstein's and Maxwell's equations. This clash is avoided in the minimal five--dimensional supergravity, which keeps the Einstein's equations from the Einstein--Maxwell theory but modifies the equations for the gauge field by adding a topological term to the action. A close five--dimensional counterpart of the Kerr-Newman solution in this theory was found in \cite{BH5d}. To generalize this construction to higher dimensions, we consider the system governed by the action
\bea\label{EinstAct}
S=\frac{1}{16\pi}\int d^d x\sqrt{-g}\left[R-\frac{1}{4}F_{\mu\nu}F^{\mu\nu}+g^2(d-1)(d-2)\right] +
S_{CS}\,.
\eea
The solution of \cite{BH5d} was obtained for such action with $d=5$ and 
\bea\label{CS5d}
S^{(d=5)}_{CS}=\frac{1}{16\pi}\frac{1}{3\sqrt{3}}\int A\wedge F\wedge F\,.
\eea
The metric and the gauge field for this case are summarized in the Appendix \ref{App5d}. 
In this article we will start with the action (\ref{EinstAct}) with undetermined Chern--Simons part and focus on the corresponding Einstein's equations 
\bea\label{Einst}
R_{\mu\nu}-\frac{1}{2}g_{\mu\nu}R=\frac{1}{2}\left[F_{\mu\alpha}{F_\nu}^\alpha-\frac{1}{4}g_{\mu\nu}F_{\alpha\beta}F^{\alpha\beta}\right]+\frac{(d-1)(d-2)}{2} g^2 g_{\mu\nu}\,.
\eea
After solving these equations for {\it both} the metric and the gauge field, we will propose the Chern--Simons term $S_{CS}$ which ensures that equations for the gauge field are satisfied as well. 

Inspired by the charged five--dimensional solution \cite{BH5d} reviewed in the Appendix \ref{App5d} and by the neutral rotating black holes in arbitrary dimensions \cite{MP,GLPPw1,GLPPw2} reviewed in the Appendix \ref{AppFrames}, we consider an ansatz for the metric and the gauge field in $d=2s+3$ dimensions\footnote{To simplify various formulas appearing in this article, 
we will interchangeably parameters $s$ and $n$ related to the number of dimensions $d$ by $d=2n+1=2s+3$.}
\bea\label{FinalBH}
ds^2&=&-\frac{\cal G}{\Xi}dt^2+\frac{\rho^2}{\Xi}ds_{\mu\phi}^2+\frac{r^2f_1}{FR\Xi^2}(dt+\omega)^2+
\frac{2qr^2f_2}{FR\Xi}\omega(dt+\omega)+
\frac{FRdr^2}{R{\cal G}-Mr^2+q f_3}\nn
A&=&\frac{qf_4}{\rho^{2s}}(dt+\omega),\quad \omega=\frac{a}{\Xi}\sum \mu_i^2 d\phi_i,\quad
ds_{\mu\phi}^2=\sum_{i=1}^{s+1}(d\mu_i^2+\mu_i^2 d\phi_i^2).
\eea
Here $(f_1,f_2,f_3,f_4)$ are undetermined functions of the radial coordinate $r$, while the other ingredients are given by
\bea\label{FinalBHp2}
{\cal G}=1+(gr)^2,\quad R=\rho^{2k},\quad 
F=\frac{r^2}{\rho^2},\quad \rho^2=r^2+a^2,\quad \Xi=1-(a g)^2.
\eea
Coordinates $\mu_i$ are subject to the constraint $\sum \mu_i^2=1$. 
Substitution of the ansatz (\ref{FinalBH}) into the Einstein's equations (\ref{Einst})
determines functions $f_j$:
\bea\label{FinalBH2}
f_1=M-\frac{sq^2}{\rho^{2s}},\quad f_2=1,\quad 
f_3=2a^2 {\cal G}+q \frac{sr^2+a^2}{\rho^{2s}},\quad
f_4=\frac{1}{\Xi}\sqrt{2s+1}\,.
\eea
Configuration (\ref{FinalBH}) does not satisfy the Maxwell's equations in the vacuum since the divergence of the Maxwell tensor gives\footnote{Here and below, $\beta$ is a numerical coefficient which depends on the number of dimensions, and its value will not be important for our discussion. For example, $\beta=(\frac{1}{3},1,2)$ for $d=(5,7,9)$.}	
\bea\label{MaxwCurr}
\nabla_\mu F^{\mu\nu}=j^\nu,\quad j^\nu\d_\nu=
\frac{24\beta\sqrt{d-2} aq^2}{\rho^{4(s+1)}}
\left[\sum \d_{\phi_k}-a\d_t\right].
\eea
In the five--dimensional case, the current is recovered by varying the Chern--Simons action (\ref{CS5d}) with respect to $A$. A standard extension of the action (\ref{CS5d}) to higher dimensions,
\bea\label{CSstand}
S^{(d=2n+1)}_{CS}=\la \int A\wedge F^n\,,
\eea
leads to
\bea
\frac{\delta S^{(2n+1)}_{CS}}{\delta A_\mu}\d_\mu=(n+1)\la \left[\star(F\wedge \dots \wedge F)\right]^\mu\d_\mu\propto 
\frac{a^{n-1}q^n}{\rho^{2n^2}}\left[\sum \d_{\phi_k}-a\d_t\right]\,,\nonumber
\eea
which matches (\ref{MaxwCurr}) only for $n=2$.
Therefore, the standard Chern--Simons term (\ref{CSstand}) cannot be used to generate equations (\ref{MaxwCurr}) beyond five dimensions, so we have to introduce some additional auxiliary fields.
Specifically, we consider the action
\bea\label{EinstActAux}
S=\frac{1}{16\pi}
\int d^d x\sqrt{-g}\left[R-\frac{1}{4}F_{\mu\nu}F^{\mu\nu}+g^2(d-1)(d-2)\right] +
\frac{1}{16\pi}\int A\wedge G^{(n)}\wedge {\bar G}^{(n)}
\eea
with exact forms $G^{(n)}=dB$ and ${\bar G}^{(n)}=d{\bar B}$. Variation with respect to the metric reproduces the Einstein's equations (\ref{Einst}), while the equations of motion for $A$, $B$, and ${\bar B}$ are
\bea\label{CSeqn3}
d\star F=G^{(n)}\wedge {\bar G}^{(n)},\quad
F\wedge G^{(n)}=0,\quad F\wedge {\bar G}^{(n)}=0.
\eea
To ensure that the first equation reproduces (\ref{MaxwCurr}), we begin with 
evaluating various ingredients appearing in (\ref{MaxwCurr}):
\bea
F&=&\frac{qf_4}{\rho^{2n}}\left[-\frac{2nr}{\rho^2}dr(dt+\omega)+d\omega\right],\quad
d\omega=\sum_j^{n-1} d\mu^2_j (d\phi_j-d\phi_n),
\nn
\star j&=&\frac{24\beta\sqrt{d-2} aq^2}{\rho^{4(n+1)}}r \rho^{d-3}dr\left[\prod \mu_j d\mu_j\right]
\left[a d\phi_{1\dots n}+\sum_{j=1}^n dt\wedge \Omega_j\right].
\eea
Here we defined a set of $n$ convenient $(n-1)$--forms:
\bea
\Omega_j=(-1)^{j+1} d\phi_1\wedge \dots \wedge 
d\phi_{j-1}\wedge d\phi_{j+1}\dots d\phi_n\,.
\eea
Defining two locally--exact 3--forms\footnote{Note that ${\bar G}$ becomes singular when $\mu_j=0$.}
\bea\label{AuxFields}
G^{(n)}=\left[\prod \mu_j d\mu_j\right](dt+\omega),\quad
{\bar G}^{(n)}=df(r)\wedge \sum_{j=1}^n \Omega_j\,,
\eea
and computing the 
products\footnote{We used the relations 
$(d\phi_j-d\phi_k)\wedge(\Omega_j-\Omega_k)=
d\phi_j\wedge\Omega_j+d\phi_k\wedge\Omega_k=0$ and 
$\omega\sum \Omega_j=a d\phi_{12\dots n}$.}
\bea
F\wedge G^{(n)}=0,\ F\wedge {\bar G}^{(n)}=0,\
G^{(n)}\wedge {\bar G}^{(n)}=-\left[\prod \mu_j d\mu_j\right]df(r)
\left[a d\phi_{1\dots n} +dt\sum_{j=1}^n \Omega_j\right],\nonumber
\eea
we conclude that the last two equations in (\ref{CSeqn3}) are satisfied. The first equation in (\ref{CSeqn3}) reduces to (\ref{MaxwCurr}) provided that 
function $f$ is defined by
\bea
f=(-1)^n \int \frac{24\beta\sqrt{d-2} aq^2}{\rho^{4(n+1)}}r \rho^{d-3}dr\,.
\eea
The last term in (\ref{EinstActAux}) represents the Chern--Simons piece of the action advertised in (\ref{EinstAct}). This concludes our discussion of  the geometry (\ref{FinalBH})--(\ref{FinalBH2}) as the solution of equations of motion coming from the action (\ref{EinstActAux}). To reiterate our logic, we started with the ansatz (\ref{FinalBH}) and determined the functions (\ref{FinalBH2}) by solving the Einstein's equations (\ref{Einst}). Then the Chern--Simons couplings and auxiliary fields were fixed by requiring that equations (\ref{CSeqn3}) reproduce the generalized Maxwell's equations (\ref{MaxwCurr}).

\bigskip

One can try to extend the solution (\ref{FinalBH})--(\ref{FinalBH2}) beyond equal rotations, but it appears that this requires significant modifications of the ansatz. Here we comment only on perturbative corrections to the neutral Myers--Perry solution in the leading power in $q$. Since the gauge potential is proportional to the charge, it does to contribute to the right hand side of the Einstein's equations (\ref{Einst}) in the linear order in $q$. Then motivated by (\ref{FinalBH}), we consider an ansatz\footnote{This expression, as well as the solution (\ref{AnstzArbRot2}) can be easily extended to linear corrections to neutral solutions with 
$g\ne 0$, but such extension of (\ref{AnstzArbRot}) is rather complicated, and it obscures our main statement.}
\bea\label{AnstzArbRot}
ds^2&=&-dt^2+\frac{Mr^2}{FR}\left(dt+\omega\right)^2
+ds_\phi^2+ds_X^2+\frac{FRdr^2}{R-Mr^2+qf_3}+
\frac{qr^2}{FR}\nu\left(dt+\omega\right),\nn
ds_X^2&=&\sum (r^2+a_i^2)d\mu_i^2,\quad 
ds_\phi^2=\sum (r^2+a_i^2)(\mu_id\phi_i)^2,\quad \omega=\sum a_i\mu_i^2d\phi_i,
\eea
where $f_3$ is an undetermined function, and $\nu$ is a one--form with undetermined coefficients $g_i$
\bea
\nu=g_0 dt+\sum g_i d\phi_i\,.
\eea
Requring the ansatz (\ref{AnstzArbRot}) to satisfy the {\it vacuum} Einsten's equation in the linear order in $q$ we find a two--parameter family of solutions
\bea\label{AnstzArbRotP2}
\nu=\alpha(dt+\omega)+\beta\sum\frac{\mu_k^2 d\phi_k}{a_k},\quad 
f_3=-\alpha r^2+\beta.
\eea
Parameter $\alpha$ corresponds to a shift in mass, while $\beta$ is a candidate for the charge--dependent correction in the metric. In particular, solution (\ref{FinalBH}) has $\beta=2a^2$, and the five--dimensional geometry constructed in \cite{BH5d} has $\beta=2a_1a_2$. Maxwell's equations at the linear order in $q$ are solved by 
\bea\label{AnstzArbRot2}
A&=&\frac{qr^2}{FR}(dt+\omega).
\eea
Unfortunately, for unequal rotations in $d>5$, we were not able to extend the geometry (\ref{AnstzArbRot})--(\ref{AnstzArbRot2}) to the full charged solution at all orders in $q$. 

\bigskip

To summarize, we have demonstrated that the geometry (\ref{FinalBH})--(\ref{FinalBH2}) solves equations of motion coming from the action (\ref{EinstActAux}). The Einstein's equations (\ref{Einst}) have sources only from the Maxwell terms, just as they do for the Kerr--Newman metric in four dimensions, while the generalized Maxwell's equations (\ref{CSeqn3}) contain sources from the auxiliary fields $(G^{(n)},{\bar G}^{(n)})$. In the five--dimensional case, the geometry (\ref{FinalBH})--(\ref{FinalBH2})  reduces to the solution found in \cite{BH5d}. 
Interestingly, in contrast to the four--dimensional Kerr--Newman case, where the metric contains only even powers of charge $q$, the solution (\ref{FinalBH}) has both even and odd powers. This phenomenon already appeared in the five dimensional case studied in \cite{BH5d}.

\section{Physical parameters of the solution}
\label{SecProperty}

In the previous section we constructed the charged rotating geometry (\ref{FinalBH})--(\ref{FinalBH2})  in all odd dimensions. Let us now analyze various physical properties of this solution. 
\begin{itemize}
\item The horizon is located at $r=r_+$, where $r_+$ is the largest positive root of
\bea
\Delta=0,\quad \Delta\equiv\frac{FR}{g_{rr}}=(R+2a^2q){\cal G}-Mr^2+q^2 \frac{sr^2+a^2}{\rho^{2s}}.
\eea
\item The entropy of the black hole is
\bea
S=\frac{(r_+^2+a^2)^n+qa^2}{4\,\Xi^n r_+}\Omega_{d-2}\,,
\eea
where $\Omega_{p}$ is the volume of the $p$--dimensional sphere.
\item The angular velocities on the horizon, $W$, are determined by requiring the Killing vector
\bea\label{AngVelE1}
\ell=\frac{\d}{\d t}-W\sum \frac{\d}{\d\phi_k}
\eea
to become null at the horizon. This gives
\bea\label{AngVelE2}
W=a\frac{(1+g^2r_+^2)(r_+^2+a^2)^{n-1}+q}{(r_+^2+a^2)^n+qa^2}\,.
\eea
\item To determine the temperature, we recall the relation
\bea
T=\frac{\kappa}{2\pi}\,,
\eea
as well as the definition of surface gravity $\kappa$,
\bea\label{KappaDef}
\ell^\mu\nabla_\mu \ell^\nu=\kappa \ell^\nu\,.
\eea
Using (\ref{AngVelE1})--(\ref{AngVelE2}), we find
\bea
\kappa=\frac{1}{2}\frac{r_+^2}{(r_+^2+a^2)}\left[1+\frac{qa^2}{(r_+^2+a^2)^n}\right]^{-1}
\lim_{r\rightarrow r_+}\left\{\d_r\frac{1}{g_{rr}}\right\}\,.
\eea
The relevant calculation is performed in the Appendix \ref{AppKappa}. The explicit expression for $\kappa$ in terms of $(r_+,q,M,a,g)$ is not very illuminating.
\item The angular momenta are
\bea
J_k=\frac{1}{16\pi}\int_{S^{d-2}}\star dK_k=\frac{na}{24\pi}\frac{M+q\Xi}{\Xi^{n+1}}
\Omega_{d-2}\,.
\eea
Here $K=\frac{\d}{\d\phi_k}$. As expected, the black hole has $n$ equal angular momenta. 
\item The electric charge of the solution is 
\bea
Q=\frac{s\sqrt{d-2}q}{16\pi\Xi^{n}}\Omega_{d-2}\,.
\eea
\item Mass of the solution can obtained by integrating the relation
\bea
dE=T dS+W_k dJ_k+\Phi dQ.
\eea
The resulting expression for $E$ is cumbersome and not very illuminating.
\end{itemize}

\section{Separation of the Klein--Gordon equation}
\label{SecKG}

In this section we will analyze excitations of the geometry (\ref{FinalBH}). First, we recall that in the neutral case, this solution has a family of Killing--Yano tensors which guarantee separation of variables in equations for scalar and spinor fields \cite{KytMpOrgW1,KytMpOrgW2,KytMpOrgW3,KytMpOrgW4,KytMpOrgW5,
KytMpMoreW1,KytMpMoreW2}\footnote{In the absence of the cosmological parameter $g$, separation 
of variables in the backgrounds of higher dimensional black holes was explored in \cite{LarsenW1,LarsenW2,LarsenW3,LarsenW4,LarsenW5}}. The equations for Maxwell and higher form bosonic fields are also separable \cite{MPmaxwOL,MPmaxwFKw1,MPmaxwFKw2,MPmaxwFKw3,MPforms}. As we will show below, introduction of charge destroys the Killing--Yano tensors, but the equation for the scalar field still remains fully integrable. 

We begin with recalling that equations for excitations of the neutral solution with multiple rotations become separable in ellipsioidal coordinates \cite{KytMpOrgW1,KytMpOrgW2,KytMpOrgW3,KytMpOrgW4,KytMpOrgW5}. In the Appendix \ref{AppFrames} we take the degenerate limit of these coordinates which is necessary for describing equal rotations. As demonstrated in that appendix, such degenerate coordinates $(r,y_1,\dots,y_{n-1})$ are given by
solving relations
\bea\label{DefEllipMain}
\mu_i^2&=&\frac{1}{{\bar c}_i^2}\prod_{k=1}^{n-1} (b_i^2-y_k),\quad
{\bar c}_i^2=\prod_{k\ne i}(b_i^2-b_k^2).
\eea
Here $(b_1,\dots,b_n)$ are arbitrary spurious parameters, and the coordinate ranges are
\bea
b_1<y_1<\dots<y_{n-1}<b_n\,.
\eea
After introduction of coordinates $(r,y_1,\dots,y_{n-1})$, the matrix inverse to the metric (\ref{FinalBH}) can be written as
\bea
g^{\mu\nu}\d_\mu\d_\nu&=&\frac{1}{FR}\left[-({\tilde e}_t)^2+({\tilde e}_r)^2\right]+
\sum \frac{1}{{\bar d}_i(r^2+a^2)}\left[({\tilde e}_i)^2+({\tilde e}_{y_i})^2\right]+
\frac{1}{r^2}({\tilde e}_\psi)^2\nn
&=&\frac{1}{FR}{\tilde g}_r^{\mu\nu}\d_\mu\d_\nu+
\sum \frac{1}{{\bar d}_i(r^2+a^2)}{\tilde g}_i^{\mu\nu}\d_\mu\d_\nu+
\frac{1}{r^2}{\tilde g}_\psi^{\mu\nu}\d_\mu\d_\nu\,.
\eea
Here we introduced separable reduced frames:
\bea\label{AllFramesAAmain}
{\tilde e}_t&=&-\frac{R}{\sqrt{\Delta}}\left[ \d_t
-\frac{a{\cal G}}{\rho^2}\d_{\psi}+\frac{qa}{\rho^{d+1}}(a\d_t-\d_\psi)\right],\quad 
{\tilde e}_r=\sqrt{{\Delta}}\d_r,\nn
{\tilde e}_i&=&\sum_k\frac{\sqrt{H_i\Xi}}{b_k^2-y_i}\d_{\phi_k},\quad 
{\tilde e}_{y_i}=2\sqrt{H_i\Xi}\d_{y_i},\quad
{\tilde e}_\psi=-a\left[\d_t-\sum_k\frac{1}{a}\d_{\phi_k}
\right],
\eea
and defined convenient combinations
\bea
{\bar d}_i=\prod_{k\ne i}(y_k-y_i),\quad {\bar H}_i=\prod_k(b_k^2-y_i).
\eea
The neutral version of these frames, (\ref{AllFramesMPGlim}), is introduced in the Appendix \ref{AppFrames}, and direct calculations show that charge appears only in ${\tilde e}_t$ and 
${\tilde e}_r$.
Note that the components $({\tilde e}^\mu_t,{\tilde e}^\mu_r)$  depend only on $r$, the components $({\tilde e}^\mu_i,{\tilde e}^\mu_{y_i})$ depend only on $y_i$, and the 
components of ${\tilde e}^\mu_\psi$ are constant. To proceed we will also need the expression for the determinant of the metric (\ref{FinalBH}) after it is rewritten in the degenerate ellipsoidal coordinates:
\bea
\sqrt{-g}=\frac{1}{2^n}\frac{FR}{r}\left[\prod \frac{{\bar d}_i}{{\bar c}_i^2}\right]^{1/2}\,.
\eea
Only the first factor in this expression depends on the radial coordinate.

Let us analyze the Klein--Gordon equation for the geometry (\ref{FinalBH}) following the general procedure introduced in \cite{MPmaxwOL}. Starting with a general equation for a charged scalar field,
\bea\label{KGgen}
(\nabla_\mu-ieA_\mu)(\nabla^\mu-ieA^\mu)\Psi-\mu^2\Psi=0,
\eea 
and using the expressions for the inverse metric and the determinant, we find\footnote{Recall that for equal rotations, $FR=r^2 \rho^{2(k-1)}$ depends only on the radial coordinate.}
\bea\label{KGpde}
&&\frac{\left[\prod{\bar d}_i\right]}{FR}\left\{r\d_\mu\left[\frac{1}{r}{\tilde e}_r^{\mu}{\tilde e}_r^{\mu}\d_\nu\Psi\right]-\left({\tilde e}^\nu_t\d_\nu-\frac{ieR}{\sqrt{\Delta}}\frac{qr^2f_4}{\rho^{2(n+1)}}\right)^2\Psi\right\}\\
&&\qquad+
\sum \frac{\left[{\prod {\bar d}_k}\right]}{{\bar d}_i(r^2+a^2)}
\d_\mu\left[{\tilde g}_i^{\mu\nu}\d_\nu\Psi\right]
+
\frac{\left[{\prod {\bar d}_k}\right]}{r^2}
\d_\mu\left[{\tilde g}_\psi^{\mu\nu}\d_\nu\Psi\right]-\mu^2\left[{\prod {\bar d}_k}\right]\Psi=0.
\nonumber
\eea
Some technical details involved in deriving this equation are presented in the Appendix \ref{AppAfield}. After imposing a separable ansatz
\bea\label{SeparAnstzPsi}
\Psi=\Phi(r)\left[\prod X_k (y_k)\right]E,\quad E\equiv \exp[i\sum_{k=0}^n n_k \phi_k],\quad \phi_0=t,
\eea
we can schematically rewrite equation (\ref{KGpde}) as
\bea
\frac{1}{X_1}\frac{\left[{\prod {\bar d}_k}\right]}{{\bar d}_1(r^2+a^2)}
\d_\mu\left[{\tilde g}_1^{\mu\nu}\d_\nu (X_1E)\right]+\dots=0,
\eea
where the omitted terms form a polynomial of degree $n-1$ in $y_1$ while having a complicated dependence on $(r,y_2,\dots,y_n)$. Consistency of separation implies that 
\bea\label{SeprPartX}
\d_\mu\left[{\tilde g}_i^{\mu\nu}\d_\nu (X_iE)\right]=-P^{(i)}_{n-1}[y_i](X_iE)
\eea
for $i=1$ and similar relations for other values of $i$. Here $P^{(i)}_{n-1}$ are $n$ independent polynomials of degree $(n-1)$. Furthermore, since ${\tilde g}_\psi^{\mu\nu}$ has indices only in cyclic directions, there is a simple relation
\bea\label{SeprPartPsi}
\d_\mu\left[{\tilde g}_\psi^{\mu\nu}\d_\nu\Psi\right]=-{\tilde g}_\psi^{\mu\nu}
n_\mu n_\nu\Psi,
\eea
where the vector $n_\mu$  has only cyclic components defined by (\ref{SeparAnstzPsi}). Substituting relations (\ref{SeprPartX}) and (\ref{SeprPartPsi}) into (\ref{KGpde}), we find
\bea\label{KGode}
\frac{1}{FR\Phi}\left\{r\d_r\left[\frac{\Delta}{r}\d_r \Phi\right]+\left({\tilde e}^\nu_t n_\nu-e{\hat A}\right)^2\right\}
-
\sum \frac{P^{(i)}_{n-1}[y_i]}{{\bar d}_i(r^2+a^2)}
-
\frac{1}{r^2}
{\tilde g}_\psi^{\mu\nu}n_\mu n_\nu-\mu^2=0.
\eea
To write this and subsequent equations in a compact form, we introduced a new function of the radial coordinate
\bea
{\hat A}=\frac{R}{\sqrt{\Delta}}\frac{qr^2f_4}{\rho^{2(n+1)}}\,.
\eea
Equation (\ref{KGode}) has apparent singularities when two $y$ coordinates approach each other, and by taking the $y_1\rightarrow y_2$ limit in (\ref{KGode}), we conclude that 
$P^{(1)}_{n-1}[y]=P^{(2)}_{n-1}[y]$. Similar arguments imply that  
\bea
P^{(1)}_{n-1}[y]=\dots =P^{(n)}_{n-1}[y]\equiv P_{n-1}[y]
\eea
Writing
\bea\label{PolynCn}
 P_{n-1}[y]=\sum_{p=0}^{n-1}c_p y^p,
\eea
we arrive at the final form of (\ref{SeprPartX}) and  (\ref{KGode}):
\bea\label{KGodeSyst}
&&\frac{1}{FR}\left\{r\d_r\left[\frac{\Delta}{r}\d_r \Phi\right]+\left({\tilde e}^\nu_t n_\nu-e{\hat A}\right)^2\Phi\right\}
-
\frac{c_{n-1}\Phi}{(r^2+a^2)}
-
\frac{1}{r^2}
{\tilde g}_\psi^{\mu\nu}n_\mu n_\nu\Phi-\mu^2\Phi=0.\nn
&&4\d_{y_i}\left[{H_i}\d_{y_i}X_i\right]-
H_i\left[\sum_k\frac{n_k}{(b_k^2-y_i)^2}\right]^2X_i=-\frac{P_{n-1}[y_i]}{\Xi}X_i
\eea
To summarize, we have demonstrated full separability of the Klein--Gordon equation  (\ref{KGgen}) by showing that it reduces to the system of ordinary differential equations (\ref{KGodeSyst}) with $n$ separation constants $(c_0,\dots c_{n-1})$ appearing in (\ref{PolynCn}). Note that only one of these constants, $c_{n-1}$, enters the radial equation. Furthermore, only the radial equation contains the parameters $(M,a,g,q)$ characterizing mass, rotation, cosmological constant, and charge of the solution. The second equation in (\ref{KGodeSyst}) describes the spherical harmonics in an unusual coordinate system.   On the other hand, the spurious parameters $(b_1,\dots b_n)$ don't enter the radial equation. This situation is in sharp contrast with the case of unequal rotations, where already in the neutral case all rotation parameters $a_k$ entered all ordinary differential equations.

In this section we have focused on separating the Klein--Gordon equation in the geometry (\ref{FinalBH})--(\ref{FinalBH2}). As reviewed in the Appendix \ref{AppFrames}, in the $q=0$ case, the spinor equations are also separable due to existence of a family of the Killing--Yano tensors (\ref{ManyKYT})--(\ref{ManyKYTh}). It is natural to ask whether such tensors would persist for the charged geometry (\ref{FinalBH})--(\ref{FinalBH2}).  By starting with one of the tensors in (\ref{ManyKYT}) and perturbing it by $q$, one can demonstrate that all such Killing--Yano tensors disappear. We have verified this explicitly in five and seven dimensions.

\section{Discussion}

In this article we constructed charged rotating solutions of the Einstein--Maxwell theory supplemented by topological couplings with auxiliary fields in all odd dimensions. Getting inspiration from the known solutions of the minimal five--dimensional supergravity, we imposed an ansatz for the metric and the gauge field and solved the relevant Einstein's equations. Then equations for the gauge field led to the unique Chern--Simons couplings in the action. We also explored various physical properties of the new solutions, in particular, we demonstrated that the Klein--Gordon equation for charged particles is fully separable on the new backgrounds. 

\bigskip

In this article we focused on black holes which have equal rotation parameters in $n$ planes, and it would be very interesting to extend our results to solutions with several independent rotations. As discussed in section \ref{SecSoln}, unfortunately such an extension would require a significant modification of the ansatz, and it is not clear whether the relevant solution can be obtained in a closed analytical form. Although in this article we have focused on black holes in odd dimensions, it appears that charged geometries in even dimensions greater than four lead to the same difficulties as the case of unequal rotations. The situation might be similar to the one encountered in higher dimensional Einstein--Maxwell equations without topological couplings: while it is clear that charged rotating solutions exist, all attempts to find them analytically have failed.

\section*{Acknowledgements}

This work was supported in part by the DOE grant DE-SC0015535.

\appendix

\section{Charged solution in five dimensions}
\label{App5d}

In this appendix we briefly summarize the geometries of the CCLP charged black holes in the minimal five--dimensional supergravity \cite{BH5d}, which served as an inspiration for the construction presented in section \ref{SecSoln}, and discuss separation of variables in charged Klein--Gordon equation on the CCLP backgrounds.

\bigskip

The authors of \cite{BH5d} constructed charged rotating black holes in the minimal five--dimensional supergravity governed by the action 
\bea
S=\frac{1}{16\pi}\int d^5x\sqrt{-g}\left[R+12 g^2-\frac{1}{4}F_{\mu\nu}F^{\mu\nu}\right]+
\frac{1}{48\sqrt{3}\pi}\int F\wedge F\wedge A
\eea
The black hole geometry is
\bea\label{CCLPsoln}
ds^2&=&-\frac{\Delta_\theta{\cal G}}{\Xi_1\Xi_2}dt^2-\frac{2q}{\rho^2}\nu\Omega+
\frac{f}{\rho^4}\Omega^2+
\rho^2\left[\frac{dr^2}{\Delta_r}+\frac{d\theta^2}{\Delta_\theta}\right]+
\sum \frac{r^2+a_i^2}{\Xi_i}\mu_i^2d\phi_i^2\,,\\
A&=&\frac{\sqrt{3}q}{\rho^2}\Omega,\quad 
\Omega=\frac{\Delta_\theta dt}{\Xi_1\Xi_2}-\omega,\quad 
\omega=\sum \frac{a_i\mu_i^2}{\Xi_i}d\phi_i,\quad
\nu=\sum {a_{3-i}\mu_i^2}d\phi_i\,,\nonumber
\eea
where various scalar functions are given by
\bea
&&\hskip -0.7cm \Delta_\theta=1-\sum(ga_{3-i}\mu_i)^2,\quad
\Delta_r=\frac{R{\cal G}+q^2+2a_1a_2 q}{r^2}-2M,\quad \rho^2=r^2+\sum (a_{3-i}\mu_i)^2,\nn
&&\hskip -0.7cm f=2M\rho^2-q^2+2a_1a_2qg^2\rho^2,\quad \Xi_i=1-(ga_i)^2,\quad
R=\prod(r^2+a_i^2),\quad {\cal G}=1+(gr)^2,\nonumber
\eea
and parameters $\mu_i$ are defined by
\bea
\mu_1=\sin\theta,\quad \mu_2=\cos\theta.
\eea
For future reference we also write the expression for the determinant of the metric (\ref{CCLPsoln}):
\bea
\sqrt{-g}=\frac{\mu_1\mu_2\rho^2 r}{\Xi_1\Xi_2}\,.
\eea
For equal rotations, the geometry becomes
\bea
ds^2&=&-\frac{{\cal G}}{\Xi}dt^2-\frac{2q}{\rho^2\Xi}{\tilde\omega}{\tilde\Omega}+
\frac{f}{\rho^4\Xi^2}{\tilde\Omega}^2+
\rho^2\left[\frac{dr^2}{\Delta_r}+\frac{1}{\Xi}(d\theta^2+
\sum\mu_i^2d\phi_i^2)\right],\\
A&=&\frac{\sqrt{3}q}{\rho^2\Xi}{\tilde\Omega},\quad 
{\tilde\Omega}=dt-{\tilde\omega},\quad 
{\tilde\omega}=a\sum {\mu_i^2}d\phi_i,\quad \Xi=1-(ag)^2.
\eea
This result inspired our ansatz (\ref{FinalBH}).

\bigskip

To connect to the discussion presented in section \ref{SecKG}, we will now analyze the Klein--Gordon equation in the geometry (\ref{CCLPsoln}). For the neutral scalar field, such analysis was carried out in \cite{0508169}\footnote{Recently fluctuations of some vector and tensor fields on the CCLP background were studied as well \cite{LarsenPert}.}, and here we will extend these results to the charged field, and more importantly, we will introduce notation that can be generalized to higher dimensions discussed in section \ref{SecKG}. 

We begin with writing the metric (\ref{CCLPsoln}) in terms of convenient frames
\bea
g^{\mu\nu}\d_\mu\d_\nu&=&\frac{1}{\rho^2}\left[-({\tilde e}_t)^2+({\tilde e}_r)^2\right]+
\frac{1}{\rho^2}\left[({\tilde e}_1)^2+({\tilde e}_{\theta})^2\right]+
\frac{1}{r^2}({\tilde e}_\psi)^2\nn
&=&\frac{1}{\rho^2}{\tilde g}_r^{\mu\nu}\d_\mu\d_\nu+
\frac{1}{\rho^2}{\tilde g}_\theta^{\mu\nu}\d_\mu\d_\nu+
\frac{1}{r^2}{\tilde g}_\psi^{\mu\nu}\d_\mu\d_\nu\,.
\eea
Here we defined several objects
\bea
&&{\tilde e}_r=\sqrt{\Delta_r}\d_r,\quad 
{\tilde e}_t=\frac{R}{r^2\sqrt{\Delta_r}}\left[\d_t+\sum\frac{a_k{\cal G}}{r^2+a_k^2}\d_{\phi_k}+\frac{qa_1a_2}{R}(\d_t+\sum\frac{1}{a_k}\d_{\phi_k})\right],\nn
&&{\tilde e}_\theta=\sqrt{\Delta_\theta}\d_\theta,\quad
{\tilde e}_1=\sqrt{\frac{1}{{\Theta}}-g^2}\left[\frac{(a_1^2-a_2^2)\mu_1\mu_2}{1-g^2\Theta}\d_t+\frac{a_1\mu_2}{\mu_1}\d_{\phi_1}-
\frac{a_2\mu_1}{\mu_2}\d_{\phi_2}\right],
\\
&&{\tilde e}_\psi=\frac{a_1a_2}{\sqrt{\Theta}}\left[\d_t+\sum\frac{1}{a_k}\d_{\phi_k}\right],\quad
\Theta=(a_1\mu_2)^2+(a_2\mu_1)^2\,.\nonumber
\eea
Note that the components $({\tilde e}^\mu_t,{\tilde e}^\mu_r)$  depend only on $r$, the components $({\tilde e}^\mu_1,{\tilde e}^\mu_{\theta})$ depend only on $\theta$, and 
the ${\tilde e}^\mu_\psi$ components  are constant. Furthermore, the charge $q$ appears only in $({\tilde e}^\mu_t,{\tilde e}^\mu_r)$. These properties persist in higher dimensions, as discussed in section \ref{SecKG}. 
The frame projections of the gauge field are
\bea
{\tilde e}_t^\mu A_\mu=\frac{\sqrt{3}q}{\sqrt{\Delta_r}}\equiv {\hat A},\quad 
{\tilde e}_r^\mu A_\mu={\tilde e}_\theta^\mu A_\mu={\tilde e}_1^\mu A_\mu={\tilde e}_\psi^\mu A_\mu=0.
\eea
Substituting these results into the Klein--Gordon equation (\ref{KGgen}) and imposing a separable ansatz,
\bea
\Psi=\Phi(r)\left[\prod X_k (y_k)\right]E,\quad
E\equiv \exp[-i\omega t+i\sum_{k=1}^n n_k \phi_k],\quad \phi_0=t,
\eea
we arrive at the system of two ordinary differential equation:
\bea\label{ODE5d1}
&&\hskip -0.7cm \frac{1}{r}\frac{d}{dr}\left[r\Delta_r\frac{d\Phi}{dr}\right]-\frac{(a_1a_2)^2}{r^2}\left[\omega-\sum\frac{n_k}{a_k}\right]^2\Phi
-\mu^2r^2 \Phi\nn
&&\hskip -0.7cm \qquad +\frac{R^2}{r^4{\Delta_r}}\left[\omega-\sum\frac{a_k{\cal G}n_k}{r^2+a_k^2}+\frac{qa_1a_2}{R}(\omega-\sum\frac{n_k}{a_k})+\frac{\sqrt{3}qer^2}{R}\right]^2\Phi=
\lambda \Phi,\\
\label{ODE5d2}
&&\hskip -0.7cm\frac{1}{s_\theta c_\theta}\frac{d}{d\theta}\left[s_\theta c_\theta\Delta_\theta\frac{dX}{d\theta}\right]-\frac{(a_1a_2)^2}{\Theta}\left[\omega-\sum\frac{n_k}{a_k}\right]^2X
-\mu^2\Theta X\nn
&&\hskip -0.7cm\qquad-\frac{\Delta_\theta (s_\theta c_\theta)^2}{{\Theta}}
\left[\frac{(a_1^2-a_2^2)\omega}{\Delta_\theta}-\frac{a_1n_1}{s_\theta}+
\frac{a_2n_2}{c_\theta}\right]^2X
=-\lambda X.
\eea
For the neutral scalar (i.e., for $e=0$) these equations are equivalent to those found in \cite{0508169}. If the two rotation parameters are equal, $a_1=a_2$, then equations (\ref{ODE5d1})--(\ref{ODE5d2}) reduce to the special case of (\ref{KGodeSyst}) for $d=5$.

\section{Neutral rotating black holes and their symmetries}
\label{AppFrames}

In this appendix we review the metrics for neutral rotating black holes in arbitrary dimensions. In the absence of the cosmological constant, such solutions were constructed by Myers and Perry in \cite{MP},  
and the AdS generalization was found by Gibbons--Lu--Page--Pope  \cite{GLPPw1,GLPPw2}. As demonstrated in \cite{KytMpOrgW1,KytMpOrgW2,KytMpOrgW3,KytMpOrgW4,KytMpOrgW5}, scalar and spinor equations on the GLPP backgrounds were separable, and an important role in ensuring this property was played by special frames. In this appendix we review the construction of these frames for generic values of rotation parameters and take a singular limit to accommodate equal rotations discussed in sections \ref{SecSoln} and \ref{SecKG}.

We begin with recalling the GLPP metric with multiple rotations \cite{GLPPw1,GLPPw2}
\bea\label{GLPPmetr}
ds^2&=&-W{\cal G}dt^2+\frac{FR dr^2}{R{\cal G}-Mr^2}-
\frac{g^2}{W{\cal G}}\Big(\sum\frac{r^2+a_i^2}{\Xi_i}\mu_id\mu_i\Big)^2\nn
&&+\frac{Mr^2}{FR}\Big(dt+\sum_{i=1}^n \frac{a_i\mu_i^2 d\phi_i}{\Xi_i}\Big)^2+
\sum_{i=1}^n\frac{r^2+a_i^2}{\Xi_i}\Big(d\mu_i^2+\mu_i^2 d{\tilde\phi}_i^2\Big).
\eea
Here we defined several objects
\bea
\Xi_i=1-(g a_i)^2,\quad {\tilde\phi}_i=\phi_i+a_i g^2 \tau,\quad
W=\sum^n \frac{\mu_i^2}{\Xi_i},\quad {\cal G}=1+(g r)^2,\nn
R=\prod (r^2+a_k^2),\quad \Delta=R-Mr^2,\quad F=1-\sum\frac{(a_k\mu_k)^2}{r^2+a_k^2},\quad \sum \mu_k^2=1.
\eea
As demonstrated in \cite{KytMpOrgW1,KytMpOrgW2,KytMpOrgW3,KytMpOrgW4,KytMpOrgW5}, the geometry (\ref{GLPPmetr}) admits a Principal Conformal Killing--Yano tensor (PCKYT), which implies full separability of equations for various dynamical fields in the background (\ref{GLPPmetr}). This PCKYT can be written in a compact form in terms of some special frames. To arrive at these frames, we begin with introducing ellipsoidal coordinates $(r,x_1,\dots x_{n-1})$, where $x_k$ are defined by
\bea\label{JacDefine}
\mu_i^2=\frac{1}{c_i^2}\prod_{k=1}^{n-1} (a_i^2-x^2_k),\quad
c_i^2=\prod_{k\ne i}(a_i^2-a_k^2).
\eea
Without loss of generality, one can choose a particular order of rotation parameters $a_k$ and ranges of the $x_k$ coordinates:
\bea\label{JacCordOrd}
0<a_1<x_1<\dots<x_{n-1}<a_n.
\eea
Then the special frames for the metric (\ref{GLPPmetr})  are defined by \cite{KytMpOrgW1,KytMpOrgW2,KytMpOrgW3,KytMpOrgW4,KytMpOrgW5}
\bea\label{AllFramesMPG}
e_i&=&-\frac{1}{Q_i}\sqrt{\frac{H_i}{x^2_id_i(r^2+x^2_i)}}\left[\d_t-\sum_k\frac{a_kQ_i^2}{a_k^2-x_i^2}\d_{\phi_k}
\right],\quad e_{x_i}=Q_i\sqrt{\frac{H_i}{x_i^2 d_i(r^2+x_i^2)}}\d_{x_i},\nn
e_t&=&-\frac{1}{Q_r}{\sqrt{\frac{R^2}{FR\Delta}}}\left[ \d_t
-\sum_k\frac{a_k{\cal G}}{r^2+a_k^2}\d_{\phi_k}\right],\quad 
e_r=Q_r\sqrt{\frac{\Delta}{FR}}\d_r,\\
e_\psi&=&-\frac{\prod a_i}{r\prod x_k}\left[\d_t-\sum_k\frac{1}{a_k}\d_{\phi_k}
\right].\nonumber
\eea
Here we defined several functions by 
\bea
Q_r=\sqrt{1+(g r)^2\frac{R}{\Delta}},\quad Q_i=\sqrt{1-(g x_i)^2},\quad
d_i=\prod_{k\ne i}(x_k^2-x_i^2),\ H_i=\prod_k(a_k^2-x_i^2).
\eea
We will also need the expression for product $FR$ in terms of ellipsoidal coordinates,
\bea
FR=r^2\prod_k(r^2+x^2_k).
\eea
As demonstrated in \cite{KytMpOrgW1,KytMpOrgW2,KytMpOrgW3,KytMpOrgW4,KytMpOrgW5}, the geometry (\ref{GLPPmetr}) admits a family of Killing--Yano tensors of various ranks which can be written in a compact form
\bea\label{ManyKYT}
Y^{2-2k}=\star\left[h^k\right]
\eea
where $h$ is the PCKYT which can be written as a two--form
\bea\label{ManyKYTh}
h=r e^r\wedge e^t+\sum_i x_i e^{x_i}\wedge e^i 
\eea
The family (\ref{ManyKYT}) implies full integrability of dynamical equations for various fields in the background (\ref{GLPPmetr}), and we refer to \cite{KytMpOrgW1,KytMpOrgW2,KytMpOrgW3,KytMpOrgW4,KytMpOrgW5,KytMpMoreW1,KytMpMoreW2} for details. In this article we are interested in the degenerate limit of the frames (\ref{AllFramesMPG}) with equal rotation parameters $a_k$. 

\bigskip

For equal rotation parameters, the naive limit of (\ref{JacDefine}) leads to a singular coordinate transformation from the set of $\mu_i$ to parameters $x_k$. To cure this problem, we write
\bea
x_k^2=a_1^2+\eps y_k,\quad a_k^2=a_1^2+\eps b_k^2
\eea
and take the $\eps\rightarrow 0$ limit of various expressions while keeping $y_k$ and $b_k$ fixed. In particular, the coordinate transformation (\ref{JacDefine}) becomes
\bea\label{JacDefineLin}
\mu_i^2=\frac{1}{{\bar c}_i^2}\prod_{k=1}^{n-1} (b_i^2-y_k),\quad
{\bar c}_i^2=\prod_{k\ne i}(b_i^2-b_k^2),
\eea
and the ranges (\ref{JacCordOrd}) translate into
\bea\label{JacCordOrdLim}
0=b^2_1<y_1<\dots<y_{n-1}<b^2_n.
\eea
The limit of frames (\ref{AllFramesMPG}) is 
\bea\label{AllFramesMPGlim}
e_i&=&-\frac{1}{a}\left[\frac{{\bar H}_i\Xi}{{\bar d}_i(r^2+a^2)}\right]^{\frac{1}{2}}
\left[-\sum_k\frac{a}{b_k^2-y_i}\d_{\phi_k}
\right],\quad e_{y_i}=2\left[\frac{{\bar H}_i\Xi}{{\bar d}_i(r^2+a^2)}\right]^{\frac{1}{2}}\d_{y_i},\nn
e_t&=&-\frac{1}{Q_r}{\sqrt{\frac{R^2}{FR\Delta}}}\left[ \d_t
-\frac{a{\cal G}}{r^2+a^2}\sum_k\d_{\phi_k}\right],\quad 
e_r=Q_r\sqrt{\frac{\Delta}{FR}}\d_r,\\
e_\psi&=&-\frac{a}{r}\left[\d_t-\sum_k\frac{1}{a}\d_{\phi_k}
\right].\nonumber
\eea
These expressions contain functions obtained by rescalings several objects encountered earlier,
\bea
{\bar d}_i=\prod_{k\ne i}(y_k-y_i),\quad {\bar H}_i=\prod_k(b_k^2-y_i),\quad 
\Xi=1-(ag)^2,
\eea
as well as some functions of the radial coordinate
\bea
R=(r^2+a^2)^n, \quad FR=r^2(r^2+a^2)^{n-1},\quad Q_r=\sqrt{1+(g r)^2\frac{R}{\Delta}},
\quad \Delta=R-Mr^2.
\eea
In section \ref{SecSoln} we constructed a charged version of the geometry (\ref{GLPPmetr}) with equal rotations, and generalization of frames (\ref{AllFramesMPGlim}) is given by (\ref{AllFramesAAmain}). Direct calculation shows that none of the Killing--Yano tensors (\ref{ManyKYT}) survive in the presence of charge. 

\section{Evaluation of surface gravity}
\label{AppKappa}

In this appendix we outline the procedure for evaluating surface gravity in an arbitrary number of dimensions and apply it to the solution (\ref{FinalBH})--(\ref{FinalBH2}). The final answer given by (\ref{KappaFinal}) is used in section \ref{SecProperty}.

\bigskip

We begin with recalling the definition of the surface gravity (\ref{KappaDef}) in terms of the Killing vector (\ref{AngVelE1})--(\ref{AngVelE2}). Using the Killing equation, we can rewrite (\ref{KappaDef}) as
\bea\label{KappaDefNew}
\nabla^\nu (\ell^\mu \ell_\mu)=-2\kappa \ell^\nu\,.
\eea
Let us compute the surface gravity for a metric that has the form
\bea\label{tempNov3}
ds^2 = g_{tt}dt^2 + 2g_{at}dt d\phi^a+g_{ab}d\phi^a d\phi^b+g_{rr}dr^2+ds_\perp^2,
\eea
where the contribution $ds_\perp^2$ built from non--cyclic coordinates remains regular at the horizon. Direct evaluation of (\ref{KappaDefNew}) in the coordinate system (\ref{tempNov3}) leads to singularities, and to cure this problem one defines a new set of coordinates $(v,\psi^a)$ by
\bea
dv=dt +\frac{g_{rr}}{g}dr,\quad 
 d\psi^a=d\phi^a - g^{ab}g_{tb}\frac{g_{rr}}{g}dr,\quad 
 g\equiv\sqrt{-g_{rr}(g_{tt}-g^{ab}g_{ta}g_{tb})}.
\eea
We will assume that function $g$ goes to a finite limit at the horizon, as it happens for all known black holes. 
This leads to the Eddington--Finkelstein form of the metric
\bea
ds^2 = g_{tt}dv^2+2g dv dr + 2 g_{ta}dv d\psi^a +g_{ab}d\psi^a d\psi^b+ds_\perp^2\,,
\eea
and the Killing vector (\ref{AngVelE2}) is given by
\bea
\ell= \d_{v}-W^a \d_{\psi^a}\,,\quad \ell_\mu dx^\mu|_{r=r_+}=g dr|_{r=r_+}\,,
\eea
where $W^a$ are angular velocities at the horizon. 
The square of the Killing vector is given by
\bea\label{temp3e}
\ell^\mu \ell_\mu =g_{ab}(g^{ac}g_{tc}-W^a)(g^{bd}g_{td}-W^b)+(g_{tt} - g_{ta}g^{ab}g_{tb}).
\eea
Notice that the first term in this expression vanishes at the horizon as $(r-r_+)^2$, so its derivative with respect to the $r$ coordinate vanishes as well. 
Taking derivative of (\ref{temp3e}) at the location of the horizon and substituting the result into 
(\ref{KappaDefNew}), we find
\bea
\kappa=\lim_{r\rightarrow r_+}\left\{-\frac{1}{2g}\d_r\left[g_{tt} - g_{ta}g^{ab}g_{tb}\right]\right\}=
\lim_{r\rightarrow r_+}\left\{\frac{1}{\sqrt{g_{rr}}}
\d_r\left[g_{ta}g^{ab}g_{tb}-g_{tt}\right]^{\frac{1}{2}}\right\}\,.
\eea
To simplify this further, we define two matrices $N_{ij}$ and $n_{ab}$ by
\bea
N_{ij}dx^idx^j=g_{tt}dt^2 + 2g_{at}dt d\phi^a+g_{ab}d\phi^a d\phi^b,\quad
n_{ab}=g_{ab}
\eea
and observe that
\bea
\mbox{det}\,N=(g_{tt} - g_{ta}g^{ab}g_{tb})\,\mbox{det}\,n\,.\nonumber
\eea
The last relation follows from the standard row decomposition\footnote{A simple way to see this is to write the metric as $g_{ab}dx^adx^b  =-N^2 dt^2+n_{ab}V^a V^b$, 
$V^a=dy^a+N^a dt$.
}. Therefore, we arrive at a compact final answer
\bea
\kappa=\lim_{r\rightarrow r_+}\left\{\frac{1}{\sqrt{g_{rr}}}\d_r\left[-\frac{\mbox{det}\,N}{\mbox{det}\,n}\right]^{\frac{1}{2}}\right\}\,.
\eea
This concludes our general discussion of surface gravity in higher dimensions.

\bigskip 

As an example, we consider the Myers--Perry geometry, i.e., the metric (\ref{GLPPmetr}) 
with $g=0$. The reduced metric is
\bea
n_{\mu\nu}dx^\mu dx^\nu&=&\frac{Mr^2}{FR}\Big(\sum_{i=1}^n {a_i\mu_i^2 d\phi_i}\Big)^2+
\sum_{i=1}^n(r^2+a_i^2)\mu_i^2 d{\phi}_i^2\,.
\eea
At the location of the horizon, function $R$ becomes equal to $Mr^2$, and this leads to drastic simplifications in the determinant of the metric:
\bea
\left[\mbox{det}\,n\right]|_{r=r_+}=\left.\frac{1}{F}\right|_{r=r_+}
\left[\prod_{k=1}^n (r_+^2+a_k^2)\mu_k^2\right]\,.
\eea
For the matrix $N$ we find
\bea
\mbox{det}\,N=-\frac{R-Mr^2}{R}\left[\prod_{k=1}^n (r^2+a_k^2)\mu_k^2\right]\,.
\eea
This leads to
\bea
\kappa=\frac{1}{2}\lim_{r\rightarrow r_+}\left[\frac{1}{R}\d_r(R-Mr^2)\right]=
\left[\sum_k\frac{r_+}{r_+^2+a_k^2}-\frac{1}{r_+}\right]\,.
\eea
In the static limit, we get $\kappa=\frac{d-3}{2r_+}$.

\bigskip

For the solution (\ref{FinalBH})--(\ref{FinalBH2}), the reduced metric is
\bea
n_{\mu\nu}dx^\mu dx^\nu&=&\frac{(ar)^2(f_1+2q\Xi)}{FR\Xi^2}\Big(\sum_{i=1}^n {\mu_i^2 d\phi_i}\Big)^2+
\frac{r^2+a^2}{\Xi}\sum_{i=1}^n \mu_i^2 d{\phi}_i^2\,.
\eea 
This leads to 
\bea
\mbox{det}\, n&=&\left[1+\frac{(ar)^2(f_1+2q\Xi)}{FR\Xi(r^2+a^2)}\right]
\left[\frac{r^2+a^2}{\Xi}\right]^n
\left[\prod_{k=1}^n \mu_k^2\right]\\
&\rightarrow&\frac{1}{\Xi}\frac{r_+^2+a^2}{r_+^2}\left[1+\frac{qa^2}{(r_+^2+a^2)^n}\right]^2
\left[\frac{r_+^2+a^2}{\Xi}\right]^n
\left[\prod_{k=1}^n \mu_k^2\right]\,.
\eea
Note that 
\bea
\mbox{det}\, N=-\frac{1}{g_{rr}}\frac{r^2}{\rho^2\Xi}
\left[\frac{r^2+a^2}{\Xi}\right]^n
\left[\prod_{k=1}^n \mu_k^2\right]\,.
\eea
This leads to the final expression
\bea\label{KappaFinal}
\kappa=\frac{1}{2}\frac{r_+^2}{(r_+^2+a^2)}\left[1+\frac{qa^2}{(r_+^2+a^2)^n}\right]^{-1}
\lim_{r\rightarrow r_+}\left\{\d_r\frac{1}{g_{rr}}\right\}\,,
\eea
which is used in section \ref{SecProperty}.

\section{Frame projections of the gauge potential}
\label{AppAfield}

In this short appendix we justify the transition between equations (\ref{KGgen}) and (\ref{KGpde}). In particular, we evaluate the projections of the gauge field $A$ from (\ref{FinalBH}) onto the frames (\ref{AllFramesAAmain}). 

Starting with the expression (\ref{FinalBH}) for the gauge field,
\bea
A=\frac{qf_4}{\rho^{2n}}(dt+a\sum \mu_k^2 d\phi_k),
\eea
and taking various projections, we find
\bea\label{ArameProj}
{\tilde e}^\mu_t A_\mu&=&\frac{R}{\sqrt{\Delta}}\frac{qf_4}{\rho^{2n}}
\left[1-\frac{a^2}{\rho^2}+\frac{qa}{\rho^{d+1}}a(1-1)\right]=
\frac{R}{\sqrt{\Delta}}\frac{qr^2f_4}{\rho^{2(n+1)}}={\tilde e}^r_r
\frac{R^2}{\Delta}\frac{qr^2f_4}{\rho^{2(n+1)}}
\nn
{\tilde e}^\mu_\psi A_\mu&=&-\frac{qaf_4}{\rho^{2n}}
\left[1-1\right]=0,\quad {\tilde e}^\mu_r A_\mu=0,\quad {\tilde e}^\mu_{y_i} A_\mu=0,\\
{\tilde e}^\mu_i A_\mu&=&\sqrt{{\bar H}_i}\frac{qf_4}{\rho^{2n}}
{\cal F}^{(n)}_i,\quad 
{\cal F}^{(n)}_i\equiv \sum_k^n \frac{1}{b_k^2-y_i}\frac{a}{{\bar c}_k^2}
\prod_{s=1}^{n-1}(b_k^2-y_s)\nonumber
\eea
Let us demonstrate that ${\cal F}^{(n)}_i=0$. We will do so by applying induction:
\begin{enumerate}[1.]
\item Direct calculations show that ${\cal F}^{(0)}_i={\cal F}^{(1)}_i=0$.
\item Assuming that  ${\cal F}^{(n)}_i=0$, we evaluate
\bea
{\cal F}^{(n+1)}_1|_{y_n=b^2_{n+1}}=
\sum_k^{n} \frac{1}{b_k^2-y_1}\frac{a}{{\bar c}_k^2}
(b_k^2-b_{n+1}^2)\prod_{s=1}^{n-1}(b_k^2-y_s)={\cal F}^{(n)}_1=0.
\eea
Similar manipulations ensure that ${\cal F}^{(n+1)}_1$ vanishes when $y_n=b^2_{k}$ for $k=1\dots n$. 
\item Definition (\ref{ArameProj}) implies that ${\cal F}^{(n+1)}_1$ is a linear function of $y_n$, and since according to item 2, it vanishes at $n$ points, this polynomial must vanish identically: 
\bea\label{FoneZero}
{\cal F}^{(n+1)}_1=0.
\eea
\item Due to symmetry between coordinates $y_k$, relation (\ref{FoneZero}) implies that
${\cal F}^{(n+1)}_1=0$. This completes the proof by induction.
\end{enumerate}
To summarize, we found 
\bea
{\tilde e}^\mu_t A_\mu=
\frac{R}{\sqrt{\Delta}}\frac{qr^2f_4}{\rho^{2(n+1)}},\quad
{\tilde e}^\mu_i A_\mu=0,\quad
{\tilde e}^\mu_\psi A_\mu=0, \quad {\tilde e}^\mu_r A_\mu=0,\quad {\tilde e}^\mu_{y_i} A_\mu=0.
\eea
Substitution of this result into (\ref{KGgen}) leads to equation (\ref{KGpde}) analyzed in section \ref{SecKG}.


\begin{thebibliography}{99}
\bibitem{BHapQGw1} 
A.~Strominger and C.~Vafa,
``Microscopic origin of the Bekenstein-Hawking entropy,''
Phys. Lett. B \textbf{379}, 99-104 (1996),
arXiv:hep-th/9601029.
\bibitem{BHapQGw2} 
C.~G.~Callan and J.~M.~Maldacena,
  ``D-brane Approach to Black Hole Quantum Mechanics,''
  Nucl.\ Phys.\  B {\bf 472}, 591 (1996), arXiv:hep-th/9602043.
\bibitem{BHapQGw3} 
 G.~T.~Horowitz and A.~Strominger,
  ``Counting States of Near-Extremal Black Holes,''
  Phys.\ Rev.\ Lett.\  {\bf 77}, 2368 (1996), arXiv:hep-th/9602051.
\bibitem{BHapQGw4} 
 J.~C.~Breckenridge, D.~A.~Lowe, R.~C.~Myers, A.~W.~Peet, A.~Strominger and C.~Vafa,
  ``Macroscopic and Microscopic Entropy of Near-Extremal Spinning Black Holes,''
  Phys.\ Lett.\  B {\bf 381}, 423 (1996), arXiv:hep-th/9603078.
  %
%
\bibitem{maldW1}
J.~M.~Maldacena,
  ``The large N limit of superconformal field theories and supergravity,''
  Adv.\ Theor.\ Math.\ Phys.\  {\bf 2}, 231 (1998), arXiv:hep-th/9711200.
  %
\bibitem{maldW2}
  S.~S.~Gubser, I.~R.~Klebanov and A.~M.~Polyakov,
  ``Gauge theory correlators from non-critical string theory,''
  Phys.\ Lett.\  B {\bf 428}, 105 (1998), arXiv:hep-th/9802109.
  %
\bibitem{maldW3}
  E.~Witten,
  ``Anti-de Sitter space and holography,''
  Adv.\ Theor.\ Math.\ Phys.\  {\bf 2}, 253 (1998), arXiv:hep-th/9802150.
\bibitem{maldW4}
 O.~Aharony, S.~S.~Gubser, J.~M.~Maldacena, H.~Ooguri and Y.~Oz,
  ``Large N field theories, string theory and gravity,''
  Phys.\ Rept.\  {\bf 323}, 183 (2000), arXiv:hep-th/9905111.
%
\bibitem{AdCFbhW1}
S.~W.~Hawking, C.~J.~Hunter and M.~Taylor,
``Rotation and the AdS / CFT correspondence,''
Phys. Rev. D \textbf{59}, 064005 (1999),
arXiv:hep-th/9811056.
%
\bibitem{AdCFbhW2}
S.~W.~Hawking and H.~S.~Reall,
``Charged and rotating AdS black holes and their CFT duals,''
Phys. Rev. D \textbf{61}, 024014 (2000),
arXiv:hep-th/9908109.
%
\bibitem{AdCFbhW3}
M.~Cvetic and S.~S.~Gubser,
``Phases of R charged black holes, spinning branes and strongly coupled gauge theories,''
JHEP \textbf{04}, 024 (1999)
arXiv:hep-th/9902195.
%
\bibitem{MP}
R.~C.~Myers and M.~J.~Perry,
  ``Black Holes in Higher Dimensional Space-Times,''
  Annals Phys.\  {\bf 172}, 304 (1986).
%
\bibitem{GLPPw1}
G.~W.~Gibbons, H.~Lu, D.~N.~Page and C.~N.~Pope,
``The General Kerr-de Sitter metrics in all dimensions,''
J. Geom. Phys. \textbf{53}, 49-73 (2005),
arXiv:hep-th/0404008.
\bibitem{GLPPw2}
G.~W.~Gibbons, H.~Lu, D.~N.~Page and C.~N.~Pope,
``Rotating black holes in higher dimensions with a cosmological constant,''
Phys. Rev. Lett. \textbf{93}, 171102 (2004),
arXiv:hep-th/0409155.
%
%
\bibitem{SenW1}
A.~Sen,
  ``Strong - weak coupling duality in four-dimensional string theory,''
  Int.\ J.\ Mod.\ Phys.\ A {\bf 9}, 3707 (1994)
  hep-th/9402002.
\bibitem{SenW2}
A.~Sen,
  ``Strong - weak coupling duality in three-dimensional string theory,''
  Nucl.\ Phys.\ B {\bf 434}, 179 (1995)
  hep-th/9408083.
\bibitem{SenW3}
  A.~Sen,
  ``Black hole solutions in heterotic string theory on a torus,''
  Nucl.\ Phys.\ B {\bf 440}, 421 (1995)
  hep-th/9411187.
\bibitem{SenW4}
M.~Cvetic and D.~Youm,
  ``Dyonic BPS saturated black holes of heterotic string on a six torus,''
Phys.\ Rev.\ D {\bf 53}, 584 (1996)
  hep-th/9507090.
\bibitem{SenW5}
M.~Cvetic and A.~A.~Tseytlin,
  ``Solitonic strings and BPS saturated dyonic black holes,''
  Phys.\ Rev.\ D {\bf 53}, 5619 (1996)
  hep-th/9512031.
\bibitem{SenW6}
M.~Cvetic and D.~Youm,
  ``All the static spherically symmetric black holes of heterotic string on a six torus,''
  Nucl.\ Phys.\ B {\bf 472}, 249 (1996)
  hep-th/9512127.
%
\bibitem{GSR4slnW1}
M.~J.~Duff and J.~T.~Liu,
``Anti-de Sitter black holes in gauged N = 8 supergravity,''
Nucl. Phys. B \textbf{554}, 237-253 (1999),
arXiv:hep-th/9901149.
%
\bibitem{GSR4slnW2}
W.~A.~Sabra,
``Anti-de Sitter BPS black holes in N=2 gauged supergravity,''
Phys. Lett. B \textbf{458}, 36-42 (1999),
arXiv:hep-th/9903143.
%
\bibitem{GSR4slnW3}
D.~D.~K.~Chow,
``Single-charge rotating black holes in four-dimensional gauged supergravity,''
Class. Quant. Grav. \textbf{28}, 032001 (2011),
arXiv:1011.2202 [hep-th].
%
%
%
\bibitem{GSR5slnW1}
K.~Behrndt, M.~Cvetic and W.~A.~Sabra,
``Nonextreme black holes of five-dimensional N=2 AdS supergravity,''
Nucl. Phys. B \textbf{553}, 317-332 (1999),
arXiv:hep-th/9810227.
%
\bibitem{GSR5slnW2}
D.~Klemm and W.~A.~Sabra,
``Charged rotating black holes in 5-D Einstein-Maxwell (A)dS gravity,''
Phys. Lett. B \textbf{503}, 147-153 (2001),
arXiv:hep-th/0010200.
\bibitem{GSR5slnW3}
M.~Cvetic, H.~Lu and C.~N.~Pope,
``Charged rotating black holes in five dimensional U(1)**3 gauged N=2 supergravity,''
Phys. Rev. D \textbf{70}, 081502 (2004),
arXiv:hep-th/0407058.
\bibitem{GSR5slnW4}
Z.~W.~Chong, M.~Cvetic, H.~Lu and C.~N.~Pope,
``Five-dimensional gauged supergravity black holes with independent rotation parameters,''
Phys. Rev. D \textbf{72}, 041901 (2005),
arXiv:hep-th/0505112.
\bibitem{GSR5slnW5}
%
Z.~W.~Chong, M.~Cvetic, H.~Lu and C.~N.~Pope,
``Non-extremal rotating black holes in five-dimensional gauged supergravity,''
Phys. Lett. B \textbf{644}, 192-197 (2007),
arXiv:hep-th/0606213.
%
\bibitem{GSR5slnW6}
S.~Q.~Wu,
``General Nonextremal Rotating Charged AdS Black Holes in Five-dimensional $U(1)^3$ Gauged Supergravity: A Simple Construction Method,''
Phys. Lett. B \textbf{707}, 286-291 (2012),
arXiv:1108.4159 [hep-th].
%
%
\bibitem{GSR6slnW1}
M.~Cvetic, H.~Lu and C.~N.~Pope,
``Gauged six-dimensional supergravity from massive type IIA,''
Phys. Rev. Lett. \textbf{83}, 5226-5229 (1999),
arXiv:hep-th/9906221.
%
\bibitem{GSR6slnW2}
D.~D.~K.~Chow,
``Charged rotating black holes in six-dimensional gauged supergravity,''
Class. Quant. Grav. \textbf{27}, 065004 (2010),
arXiv:0808.2728 [hep-th].
%
%
\bibitem{GSR7slnW1}
Z.~W.~Chong, M.~Cvetic, H.~Lu and C.~N.~Pope,
``Non-extremal charged rotating black holes in seven-dimensional gauged supergravity,''
Phys. Lett. B \textbf{626}, 215-222 (2005),
arXiv:hep-th/0412094.
%
\bibitem{GSR7slnW2}
S.~Q.~Wu,
``Two-charged non-extremal rotating black holes in seven-dimensional gauged supergravity: The Single-rotation case,''
Phys. Lett. B \textbf{705}, 383-387 (2011).
arXiv:1108.4158 [hep-th].


\bibitem{10auth}
M.~Cvetic, M.~J.~Duff, P.~Hoxha, J.~T.~Liu, H.~Lu, J.~X.~Lu, R.~Martinez-Acosta, C.~N.~Pope, H.~Sati and T.~A.~Tran,
``Embedding AdS black holes in ten-dimensions and eleven-dimensions,''
Nucl. Phys. B \textbf{558}, 96-126 (1999),
arXiv:hep-th/9903214.
%
%
%
\bibitem{ByndGauSugraW1}
H.~Lu,
``Charged dilatonic ads black holes and magnetic AdS$_{D-2} \times R^{2}$ vacua,''
JHEP \textbf{09}, 112 (2013)
arXiv:1306.2386 [hep-th].
%
\bibitem{ByndGauSugraW2}
S.~Q.~Wu,
``General rotating charged Kaluza-Klein AdS black holes in higher dimensions,''
Phys. Rev. D \textbf{83}, 121502 (2011)
arXiv:1108.4157 [hep-th].
%
\bibitem{ByndGauSugraW3}
R.~Deshpande and O.~Lunin,
``Multi--charged geometries with cosmological constant,''
arXiv:2408.08254 [hep-th].
%
%
%
%
\bibitem{KNdFailW1}
A.~N.~Aliev,
``A Slowly rotating charged black hole in five dimensions,''
Mod. Phys. Lett. A \textbf{21}, 751-758 (2006),
arXiv:gr-qc/0505003.
%
\bibitem{KNdFailW2}
A.~N.~Aliev,
``Rotating black holes in higher dimensional Einstein-Maxwell gravity,''
Phys. Rev. D \textbf{74}, 024011 (2006),
arXiv:hep-th/0604207 [hep-th].
%
\bibitem{KNdFailW3}
P.~Krtous,
``Electromagnetic field in higher-dimensional black-hole spacetimes,''
Phys. Rev. D \textbf{76}, 084035 (2007),
arXiv:0707.0002 [hep-th].
%
%
\bibitem{KNdFailW4}
M.~Allahverdizadeh, J.~Kunz and F.~Navarro-Lerida,
``Extremal Charged Rotating Black Holes in Odd Dimensions,''
Phys. Rev. D \textbf{82}, 024030 (2010),
arXiv:1004.5050.
%
\bibitem{KNdFailW5}
M.~Ortaggio and A.~Srinivasan,
``Charging Kerr-Schild spacetimes in higher dimensions,''
Phys. Rev. D \textbf{110}, no.4, 4 (2024),
arXiv:2309.02900 [gr-qc].
%
\bibitem{BH5d} 
Z.~W.~Chong, M.~Cvetic, H.~Lu and C.~N.~Pope,
``General non-extremal rotating black holes in minimal five-dimensional gauged supergravity,''
Phys. Rev. Lett. \textbf{95}, 161301 (2005),
arXiv:hep-th/0506029.
%
%
\bibitem{KytMpOrgW1}
V.~P.~Frolov and D.~Kubiznak,
``Hidden Symmetries of Higher Dimensional Rotating Black Holes,''
Phys. Rev. Lett. \textbf{98}, 011101 (2007),
arXiv:gr-qc/0605058 [gr-qc].
%
\bibitem{KytMpOrgW2}
D.~Kubiznak and V.~P.~Frolov,
``Hidden Symmetry of Higher Dimensional Kerr-NUT-AdS Spacetimes,''
Class. Quant. Grav. \textbf{24}, no.3, F1-F6 (2007),
arXiv:gr-qc/0610144.
\bibitem{KytMpOrgW3}
D.~N.~Page, D.~Kubiznak, M.~Vasudevan and P.~Krtous,
``Complete integrability of geodesic motion in general Kerr-NUT-AdS spacetimes,''
Phys. Rev. Lett. \textbf{98}, 061102 (2007),
arXiv:hep-th/0611083 [hep-th].
%
\bibitem{KytMpOrgW4}
V.~P.~Frolov, P.~Krtous and D.~Kubiznak,
``Separability of Hamilton-Jacobi and Klein-Gordon Equations in General Kerr-NUT-AdS Spacetimes,''
JHEP \textbf{02}, 005 (2007),
arXiv:hep-th/0611245 [hep-th].
\bibitem{KytMpOrgW5}
P.~Krtous, D.~Kubiznak, D.~N.~Page and V.~P.~Frolov,
``Killing-Yano Tensors, Rank-2 Killing Tensors, and Conserved Quantities in Higher Dimensions,''
JHEP \textbf{02}, 004 (2007),
arXiv:hep-th/0612029 [hep-th].
\bibitem{ChL} Y.~Chervonyi and O.~Lunin,
``Killing(-Yano) Tensors in String Theory,''
JHEP \textbf{09}, 182 (2015),
arXiv:1505.06154 [hep-th].
%
%
%
\bibitem{MPmaxwOL} 
O.~Lunin,
``Maxwell\textquoteright{}s equations in the Myers-Perry geometry,''
JHEP \textbf{12}, 138 (2017),
arXiv:1708.06766 [hep-th].
\bibitem{MPmaxwFKw1} 
P.~Krtou\v{s}, V.~P.~Frolov and D.~Kubiz\v{n}\'ak,
``Separation of Maxwell equations in Kerr\textendash{}NUT\textendash{}(A)dS spacetimes,''
Nucl. Phys. B \textbf{934}, 7-38 (2018),
arXiv:1803.02485.
%
\bibitem{MPmaxwFKw2} 
V.~P.~Frolov, P.~Krtou\v{s}, D.~Kubiz\v{n}\'ak and J.~E.~Santos,
``Massive Vector Fields in Rotating Black-Hole Spacetimes: Separability and Quasinormal Modes,''
Phys. Rev. Lett. \textbf{120}, 231103 (2018),
arXiv:1804.00030 [hep-th].
%
\bibitem{MPmaxwFKw3} 
R.~Cayuso, O.~J.~C.~Dias, F.~Gray, D.~Kubiz\v{n}\'ak, A.~Margalit, J.~E.~Santos, R.~Gomes Souza and L.~Thiele,
``Massive vector fields in Kerr-Newman and Kerr-Sen black hole spacetimes,''
JHEP \textbf{04}, 159 (2020),
arXiv:1912.08224 [hep-th].
\bibitem{MPforms} O.~Lunin,
``Excitations of the Myers-Perry Black Holes,''
JHEP \textbf{10}, 030 (2019),
arXiv:1907.03820 [hep-th].
%
%
\bibitem{KubizKYT5}
D.~Kubiznak, H.~K.~Kunduri and Y.~Yasui,
``Generalized Killing-Yano equations in D=5 gauged supergravity,''
Phys. Lett. B \textbf{678}, 240-245 (2009),
arXiv:0905.0722.
%
\bibitem{KytMpMoreW1} 
V.~P.~Frolov and D.~Kubiznak,
``Higher-Dimensional Black Holes: Hidden Symmetries and Separation of Variables,''
Class. Quant. Grav. \textbf{25}, 154005 (2008),
arXiv:0802.0322 [hep-th].
%
\bibitem{KytMpMoreW2} 
P.~Krtous, V.~P.~Frolov and D.~Kubiznak,
``Hidden Symmetries of Higher Dimensional Black Holes and Uniqueness of the Kerr-NUT-(A)dS spacetime,''
Phys. Rev. D \textbf{78}, 064022 (2008),
arXiv:0804.4705 [hep-th].
%
\bibitem{LarsenW1}
M.~Cvetic and F.~Larsen,
``General rotating black holes in string theory: Grey body factors and event horizons,''
Phys. Rev. D \textbf{56}, 4994-5007 (1997),
arXiv:hep-th/9705192.
%
\bibitem{LarsenW2}
M.~Cvetic and F.~Larsen,
``Grey body factors for rotating black holes in four-dimensions,''
Nucl. Phys. B \textbf{506}, 107-120 (1997),
arXiv:hep-th/9706071.
%
\bibitem{LarsenW3}
M.~Cvetic and F.~Larsen,
``Near horizon geometry of rotating black holes in five-dimensions,''
Nucl. Phys. B \textbf{531}, 239-255 (1998),
arXiv:hep-th/9805097.
%
\bibitem{LarsenW4}
M.~Cvetic and F.~Larsen,
``Greybody Factors and Charges in Kerr/CFT,''
JHEP \textbf{09}, 088 (2009),
arXiv:0908.1136 [hep-th].
%
\bibitem{LarsenW5}
C.~Keeler and F.~Larsen,
``Separability of Black Holes in String Theory,''
JHEP \textbf{10}, 152 (2012),
arXiv:1207.5928 [hep-th].
%
\bibitem{0508169}P.~Davis, H.~K.~Kunduri and J.~Lucietti,
``Special symmetries of the charged Kerr-AdS black hole of D=5 minimal gauged supergravity,''
Phys. Lett. B \textbf{628}, 275-280 (2005),
arXiv:hep-th/0508169 [hep-th].
%
\bibitem{LarsenPert}
N.~Ezroura and F.~Larsen,
``Supergravity spectrum of AdS$_{5}$ black holes,''
JHEP \textbf{12}, 020 (2024),
arXiv:2408.11529 [hep-th].
%
%
%
%
%
\end{thebibliography}
\end{document}